\documentclass[epj]{webofc}
\usepackage[utf8]{inputenc}
\usepackage[varg]{txfonts}   
\usepackage{booktabs}
\usepackage{xcolor}
\definecolor{darkred}{rgb}{0.4,0.0,0.0}
\definecolor{darkgreen}{rgb}{0.0,0.4,0.0}
\definecolor{darkblue}{rgb}{0.0,0.0,0.4}
\usepackage[bookmarks,linktocpage,colorlinks,
    linkcolor = darkred,
    urlcolor  = darkblue,
    citecolor = darkgreen]{hyperref}
\usepackage{subcaption}
\usepackage{amsfonts}
\usepackage{dsfont}
\wocname{EPJ Web of Conferences}
\woctitle{Lattice2017}
\DeclareMathOperator{\Tr}{Tr}
\begin{document}
%
\selectlanguage{english}
\title{%
Towards leading isospin breaking effects in mesonic masses with $O(a)$ improved Wilson fermions
}
\author{%
\firstname{Andreas} \lastname{Risch}\inst{1}\fnsep\thanks{Speaker, \email{andreas.risch@uni-mainz.de}} \and
\firstname{Hartmut} \lastname{Wittig}\inst{1,2}
}
\institute{%
PRISMA Cluster of Excellence and Institut f{\"u}r Kernphysik, University of Mainz, Johann-Joachim-Becher-Weg 45, 55099 Mainz, Germany
\and
Helmholtz Institute Mainz, University of Mainz, 55099 Mainz, Germany
}
\abstract{%
We present an exploratory study of leading isospin breaking effects in mesonic masses using $O(a)$ improved Wilson fermions. Isospin symmetry is explicitly broken by distinct masses and electric charges of the up and down quarks. In order to be able to make use of existing isosymmetric QCD gauge ensembles we apply reweighting techniques. The path integral describing QCD+QED is expanded perturbatively in powers of the light quarks' mass deviations and the electromagnetic coupling. We employ QED$_{\mathrm{L}}$ as a finite volume formulation of QED.
}
\maketitle
\section{Introduction}
\label{sec_introduction}

Current lattice simulations investigating hadron physics are usually performed in the isosymmetric limit, i.e. with degenerate light quark masses, only taking QCD into account and neglecting QED effects~\cite{Aoki:2016frl}. In order to be sensitive to possible deviations from experimental determinations of many observables  like the muon anomalous magnetic moment and the masses of the light hadrons a statistical precision of $O(1\%)$ and below is required in lattice simulations~\cite{Patrignani:2016xqp}. However, at this level of precision isospin breaking effects have to be considered. In particular cases effects due to QED and the light quark mass splitting have the same order of magnitude but opposite signs~\cite{Portelli:2015wna}. Hence, for consistency both the light quark mass non-degeneracy as well as electromagnetic effects have to be included in lattice simulations~\cite{Portelli:2015wna, Patella:2017fgk}.

In this work we adopt the idea of treating isospin breaking effects perturbatively developed by the ROME123 collaboration~\cite{Sanfilippo:2011zz, deDivitiis:2011eh, Tantalo:2013maa, deDivitiis:2013xla}. We organise this work as follows: We first introduce the setup of QCD+QED simulations based on reweighting and perturbation theory, give a set of Feynman rules for the calculation of correlation functions to leading order isospin breaking with $O(a)$ improved Wilson fermions and describe the expansion of a generic mesonic 2-point function. We discuss the stochastic treatment of all-to-all photon propagators, further specify the extraction of pseudo-scalar ground state masses and finally give exploratory results for one gauge ensemble.

\section{Inclusion of perturbative isospin breaking effects by reweighting}
\label{sec_inclusion_of_isospin_breaking}

In this exploratory study we utilise existing QCD gauge ensembles created by the CLS effort~\cite{Fritzsch:2012wq}. The latter are $N_{f}=2$ ensembles with an $O(a)$ improved Wilson quark action $S^{(0)}_{\mathrm{q}}[U] = \overline{\Psi}_{\mathbf{a}}{D^{(0)}[U]^{\mathbf{a}}}_{\mathbf{b}}\Psi^{\mathbf{b}}$ including two degenerate dynamical light quarks with $m^{(0)\mathrm{u}}=m^{(0)\mathrm{d}}$ and Wilson plaquette action $S_{\mathrm{g}}^{(0)}$ on (anti-)periodic boundary conditions. The strange quark is included as a valence quark. Quark fields carry Euclidean spacetime $x$, flavour $f$, colour $c$ and spin $s$ degrees of freedom. We apply a condensed index notation, where $\mathbf{a},\mathbf{b}\equiv (xfcs)$ describe quark field indices. The metric tensor is unity if not stated otherwise. In the following we will refer to the theory described as $\mathrm{QCD}_{\mathrm{iso}}$. Quantities defined in this theory will be labelled with the superscript $(0)$. The gauge expectation value reads
\begin{align*}
\left\langle O[U] \right\rangle_{\mathrm{g}}^{(0)} &= \frac{1}{Z^{(0)}} \int DU \exp\left(-S^{(0)}_{\mathrm{g}}[U]\right)Z^{(0)}_{\mathrm{q}}[U]\,O[U] & Z^{(0)}_{\mathrm{q}}[U] &= \int D\Psi D\overline{\Psi} \exp\left(-S^{(0)}_{\mathrm{q}}[U,\Psi,\overline{\Psi}]\right),
\end{align*}
where the partition function $Z^{(0)}$ is determined by the condition $\left\langle 1 \right\rangle_{\mathrm{g}}^{(0)}=1$.

In order to be able to reuse gauge ensembles of QCD$_{\mathrm{iso}}$ for the calculation of expectation values in QCD+QED we apply reweighting techniques~\cite{Tantalo:2013maa, deDivitiis:2013xla}. We introduce the QED expectation value on a classical QCD background field with a non-compact photon action $S_{\gamma}$ and quark action $S_{\mathrm{q}}$ reading
\begin{align*}
\left\langle O[U,A,\Psi,\overline{\Psi}] \right\rangle_{\mathrm{q}\gamma} &= \frac{1}{Z_{\mathrm{q}\gamma}[U]} \int DA D\Psi D\overline{\Psi} \exp\left(-S_{\gamma}[A]-S_{\mathrm{q}}[U,A,\Psi,\overline{\Psi}]\right)\,O[U,A,\Psi,\overline{\Psi}],
\end{align*}
where again $Z_{\mathrm{q}\gamma}[U]$ is determined by $\left\langle 1 \right\rangle_{\mathrm{q}\gamma}=1$.
Consequently, expectation values in QCD+QED can be expressed as
\begin{align}
\left\langle O[U,A,\Psi,\overline{\Psi}] \right\rangle &= \frac{\left\langle R[U] \left\langle O[U,A,\Psi,\overline{\Psi}] \right\rangle_{\mathrm{q}\gamma} \right\rangle_{\mathrm{g}}^{(0)}}{\left\langle R[U] \right\rangle_{\mathrm{g}}^{(0)}} &
R\lbrack U \rbrack &=  \frac{\exp\left(-S_{\mathrm{g}}[U]\right)Z_{\mathrm{q}\gamma}[U]}{\exp\left(-S_{\mathrm{g}}^{(0)}[U]\right)Z^{(0)}_{\mathrm{q}}[U]},
\label{eq_expectation_value_by_reweighting}
\end{align}
where we have introduced the reweighting factor $R$, which is given by the ratio of weighting factors of the effective QCD gauge actions in QCD+QED and QCD$_{\mathrm{iso}}$. While we use $S^{(0)}_{\mathrm{g}}=S_{\mathrm{g}}$, one may make use of differently scaled gauge actions in order to account for a renormalisation of the strong coupling due to isospin breaking effects~\cite{Tantalo:2013maa, deDivitiis:2013xla}.

Non-compact finite volume QED suffers from zero-mode divergencies and therefore requires an infrared regularisation. Furthermore, the naive formulation is invariant under unphysical large gauge transformations. We utilise the lattice discretisation QED$_{\mathrm{L}}$~\cite{Patella:2017fgk, Hayakawa:2008an, Borsanyi:2014jba}, where the spatial zero modes do not belong to the dynamic degrees of freedom and consequently are set to zero: $\widetilde{A}^{p\mu}=0$ for $\vec{p}=0$. This procedure also removes the invariance under large gauge transformations mentioned. As we use a non-compact formulation we have to apply gauge fixing. We use the Feynman gauge. The photon action reads
\begin{align}
S_{\gamma}[A] &= \frac{1}{2}A_{\mathbf{c_{2}}}{\Delta^{\mathbf{c_{2}}}}_{\mathbf{c_{1}}}A^{\mathbf{c_{1}}} & \widetilde{\Delta}^{p_{2}\mu_{2}}_{\hphantom{{p_{2}\mu}}p_{1}\mu_{1}} &= 4 \sum_{\nu} \sin^{2}\left(\frac{p_{1\nu}}{2}\right) \,\delta^{p_{2}}_{p_{1}}\delta^{\mu_{2}}_{\mu_{1}},
\label{eq_photon_action}
\end{align}
where we applied the condensed index notation $\mathbf{c}\equiv (x\mu)$ for photon field indices. With Fourier transforms ${\mathcal{F}^{p}}_{x} = {|\Lambda|}^{-1/2}\exp(-\mathrm{i}p_{\mu}x^{\mu})$ and $ {\mathcal{F}^{-1x}}_{p} = {|\Lambda|}^{-1/2}\exp(\mathrm{i}p_{\mu}x^{\mu})$ we have
${\Delta^{x_{2}\mu_{2}}}_{x_{1}\mu_{1}} = {{\mathcal{F}^{-1}}^{x_{2}}}_{p_{2}} {\widetilde{\Delta}}^{p_{2}\mu_{2}}_{\hphantom{p_{2}\mu_{2}}p_{1}\mu_{1}} {\mathcal{F}^{p_{1}}}_{x_{1}}$. In Fourier space the metric tensor reads $\widetilde{g}_{p_{2}p_{1}} = \delta^{p_{2}}_{-p_{1}}$.

The fermion action $S_{\mathrm{q}}[U,A,\Psi,\overline{\Psi}] = \overline{\Psi}_{\mathbf{a}}{D[U,A]^{\mathbf{a}}}_{\mathbf{b}}\Psi^{\mathbf{b}}$ is related to the fermion action $S^{(0)}_{\mathrm{q}}$ of the underlying isosymmetric theory. In order to introduce the electromagnetic interaction, we define gauge links $V$~\cite{Duncan:1996xy, Tantalo:2013maa, deDivitiis:2013xla} as
\begin{align}
V^{x\mu f} &= U^{x\mu}\exp\left(\mathrm{i}ee^{f}A^{x\mu}\right).
\label{eq_QCD_QED_gauge_link}
\end{align}
$S_{\mathrm{q}}$ is now built from $S^{(0)}_{\mathrm{q}}$ by the replacement of gauge links $U$ with $V$ in all terms but the Clover term. In QCD+QED we only work with the Clover term inherited from the underlying QCD$_{\mathrm{iso}}$. Additionally, the quark masses can be detuned arbitrarily, i.e. we work with non-degenerate bare light quark masses $m^{\mathrm{u}}\neq m^{\mathrm{d}}$.

In order to evaluate $\left\langle \ldots \right\rangle_{\mathrm{q}\gamma}$ and $Z_{\mathrm{q}\gamma}$ in Eq.~\ref{eq_expectation_value_by_reweighting} we treat isospin breaking effects in the framework of weak coupling perturbation theory ~\cite{Sanfilippo:2011zz, deDivitiis:2011eh, Tantalo:2013maa, deDivitiis:2013xla}. The path integral is expanded in the parameters $\Delta m^{f} = m^{f}-m^{(0)f}$ and $e^{2}$. In this work we only consider first-order contributions. The electromagnetic coupling does not renormalise to this order and hence can be fixed to $e^{2}=4\pi\alpha$. The Wilson-Dirac-Operator $D[U,A]$ of QCD+QED is expanded around the Wilson-Dirac-Operator $D^{(0)}[U]$ of QCD$_{\mathrm{iso}}$. This expansion gives us the appropriate vertices for the Feynman rules:
\begin{align*}
{D[U,A]^{\mathbf{a}}}_{\mathbf{b}} &= {D^{(0)}[U]^{\mathbf{a}}}_{\mathbf{b}} + V_{\mathrm{qq}}{{}^{\mathbf{a}}}_{\mathbf{b}}[U] + V_{\mathrm{qq\gamma}}[U]{{}^{\mathbf{a}}}_{\mathbf{bc}}\,A^{\mathbf{c}} + \frac{1}{2}V_{\mathrm{qq\gamma\gamma}}[U]{{}^{\mathbf{a}}}_{\mathbf{bc_{2}c_{1}}}\,A^{\mathbf{c_{2}}}A^{\mathbf{c_{1}}} + \ldots .
\end{align*}
We introduce a graphical representations for the vertices:
\begin{align*}
V_{\mathrm{qq}}{{}^{\mathbf{a}}}_{\mathbf{b}} &=
\begin{gathered}
\includegraphics[width=4em]{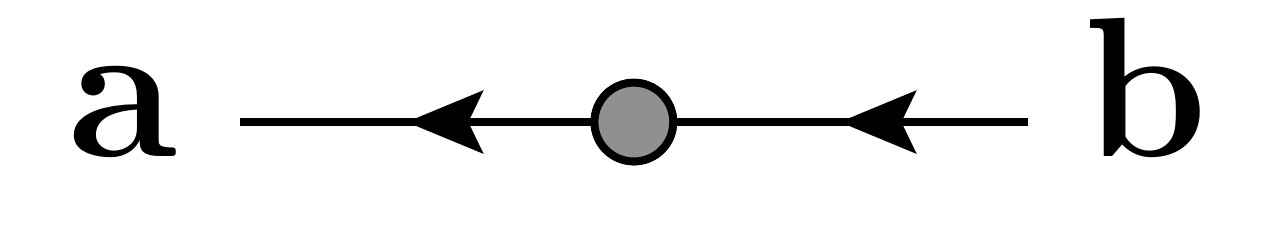}
\end{gathered}
& V_{\mathrm{qq\gamma}}{{}^{\mathbf{a}}}_{\mathbf{bc}} &=
\begin{gathered}
 \includegraphics[width=4em]{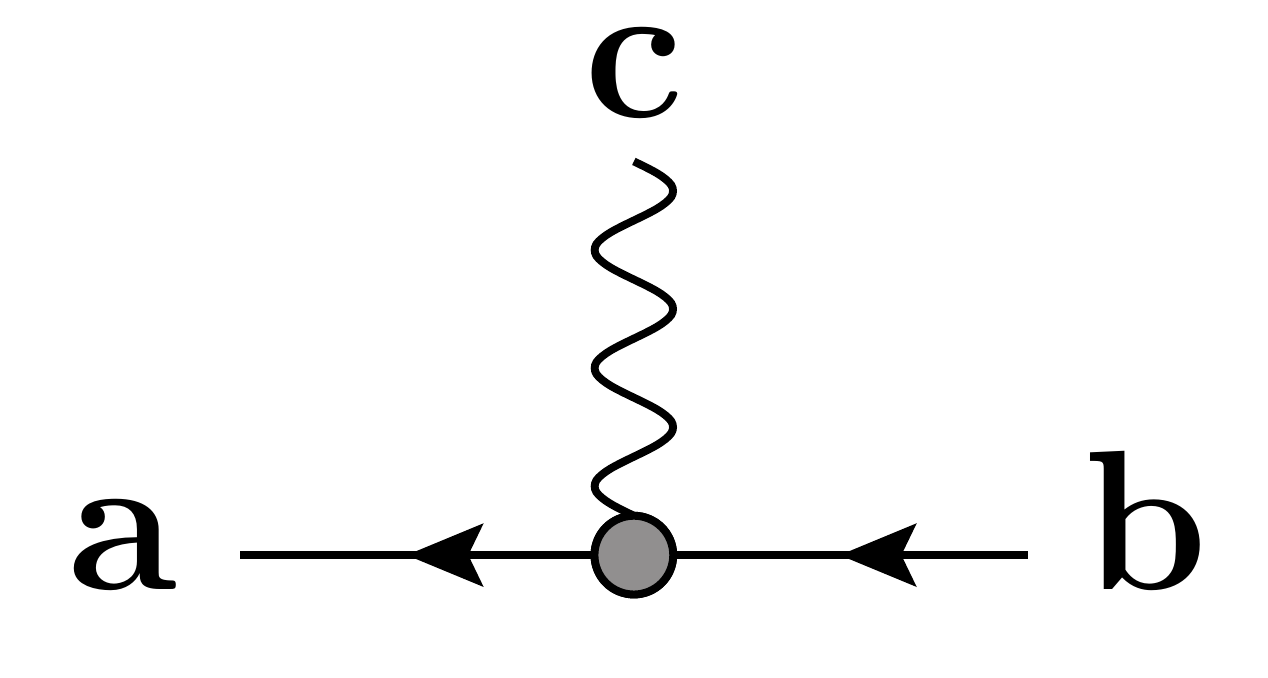}
\end{gathered}
& V_{\mathrm{qq\gamma\gamma}}{{}^{\mathbf{a}}}_{\mathbf{bc_{1}c_{2}}} &=
\begin{gathered}
\includegraphics[width=4em]{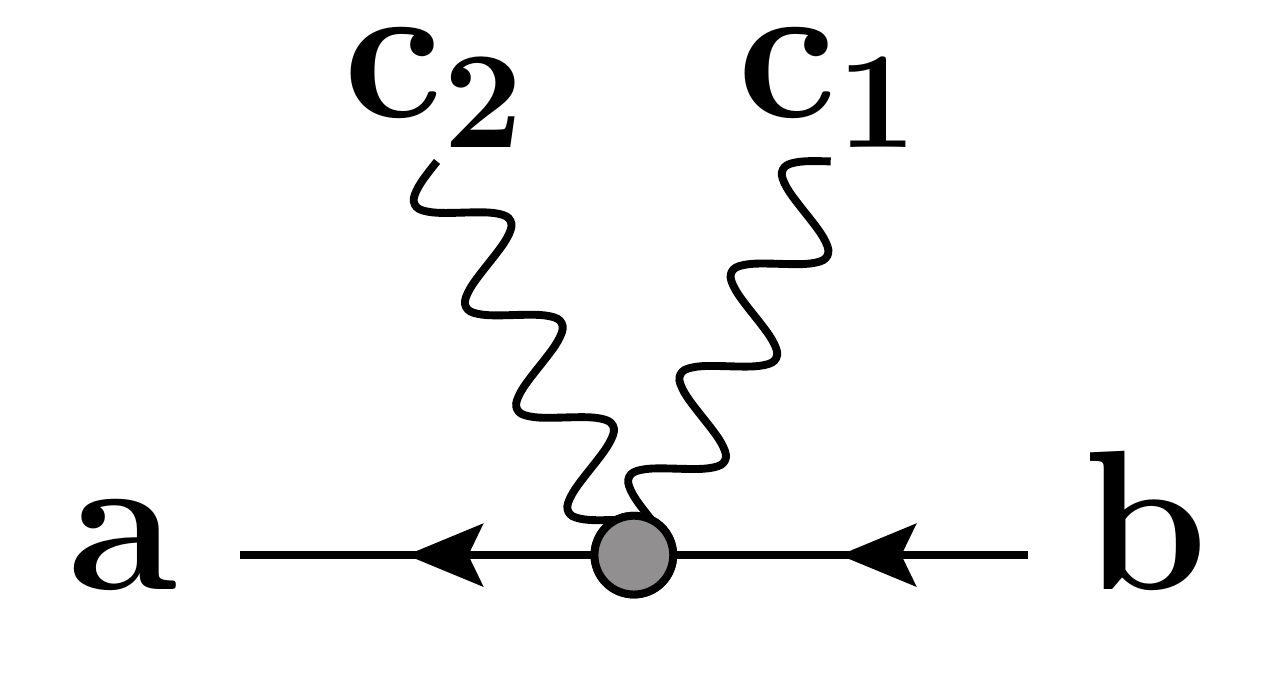}
\end{gathered}
.
\end{align*}
For convenience and later purpose the vertices are contracted with a quark and photon fields. Colour and spinor indices are suppressed. In the case of Wilson fermions the vertices are given by
\begin{align}
V_{\mathrm{qq}}{{}^{xf}}_{\mathbf{b}}\Psi^{\mathbf{b}} &= \left(m^{f}-m^{(0)f}\right)\Psi^{xf} 
\label{eq_contracted_vertices1} \\
V_{\mathrm{qq\gamma}}{{}^{xf}}_{\mathbf{bc}}\Psi^{\mathbf{b}}A^{\mathbf{c}} &= \frac{\mathrm{i}}{2}ee^{f}\sum_{\mu}\left(\left(\gamma^{\mu} - \mathds{1}\right) U^{x\mu} A^{x\mu} \Psi^{x+\hat{\mu},f} + \left(\gamma^{\mu} + \mathds{1}\right) {U^{x-\hat{\mu},\mu}}^{\dagger} A^{x-\hat{\mu},\mu} \Psi^{x-\hat{\mu},f}\right) \nonumber \\
V_{\mathrm{qq\gamma\gamma}}{{}^{xf}}_{\mathbf{bc_{2}c_{1}}}\Psi^{\mathbf{b}}A^{\mathbf{c_{2}}}A^{\mathbf{c_{1}}} &= \frac{1}{2}\left(ee^{f}\right)^{2}\sum_{\mu}\left(\left(- \gamma^{\mu} + \mathds{1}\right) U^{x\mu} \left({A^{x\mu}}\right)^{2} \Psi^{x+\hat{\mu},f} + \left(\gamma^{\mu} + \mathds{1}\right) {U^{x-\hat{\mu},\mu}}^{\dagger} \left({A^{x-\hat{\mu},\mu}}\right)^{2} \Psi^{x-\hat{\mu},f}\right). \nonumber
\end{align}
The quark propagator is extracted from QCD$_{\mathrm{iso}}$ and calculated as usual by numerical inversion on a classical QCD background field. The photon propagator in Feynman gauge can be derived from Eq.~\ref{eq_photon_action} and is given by~\cite{Boyle:2016lbc}
\begin{align}
{\Sigma^{x_{2}\mu_{2}}}_{x_{1}\mu_{1}} &= {{\mathcal{F}^{-1}}^{x_{2}}}_{p_{2}} {\widetilde{\Sigma}}^{p_{2}\mu_{2}}_{\hphantom{p_{2}\mu_{2}}p_{1}\mu_{1}} {\mathcal{F}^{p_{1}}}_{x_{1}}
& {\widetilde{\Sigma}}^{p_{2}\mu_{2}}_{\hphantom{p_{2}\mu_{2}}p_{1}\mu_{1}} &= \begin{cases}
\Big(4 \sum_{\nu} \sin^{2}\Big(\frac{p_{1\nu}}{2}\Big)\Big)^{-1} \delta^{p_{1}}_{p_{2}}\delta^{\mu_{2}}_{\mu_{1}} & \text{ for } \vec{p_{2}},\vec{p_{1}}\neq 0 \\
0 & \text{ otherwise}
\end{cases},
\label{eq_photon_propagator}
\end{align}
where we have made use of its Fourier space representation $\widetilde{\Sigma}$. The projection is due to the infrared regularisation in QED$_{\mathrm{L}}$. We use the standard graphical representations for quark and photon propagators:
\begin{align*}
S^{(0)}{{}^{\mathbf{b}}}_{\mathbf{a}}&= \includegraphics[width=4em]{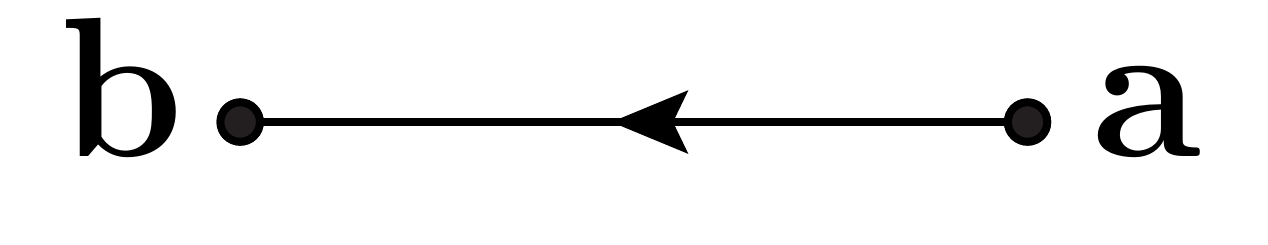} & {\Sigma^{\mathbf{c_{2}}}}_{\mathbf{c_{1}}}&= \includegraphics[width=4em]{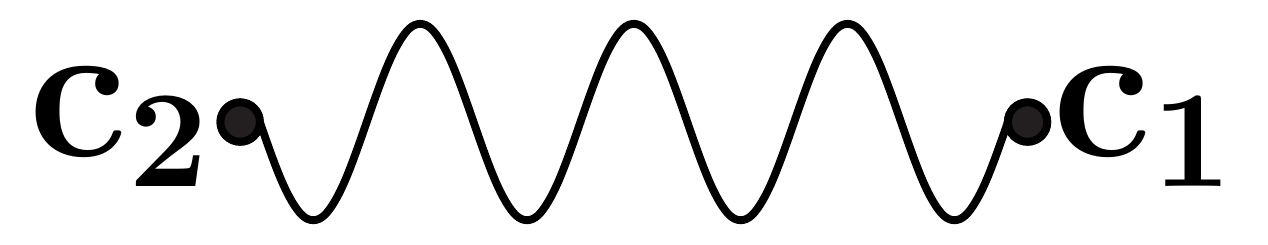}.
\end{align*}
For $\langle \ldots \rangle_{\mathrm{q}\gamma}$ only diagrams contribute in which all parts are connected to the operators inserted. For each closed fermion loop as well as for each vertex we have to multiply by factors of $-1$ in our calculation. We also have to divide by the appropriate symmetry factor of the diagram. The quark determinant $Z_{\mathrm{q}\gamma}$ is expanded in the standard fashion.

\section{Extraction of pseudo-scalar meson masses from 2-point functions}
\label{sec_meson_masses}

In the following, we consider the perturbative expansion of a generic mesonic 2-point-function  of point-like operators, i.e. operators that do not depend on the gauge links $V$ defined in Eq.~\ref{eq_QCD_QED_gauge_link}, evaluated according to Eq.~\ref{eq_expectation_value_by_reweighting}. We restrict our analysis to quark-connected diagrams and neglect vacuum contributions. In this approximation the 2-point-function reads
\begin{align}
\left\langle \overline{\Psi}_{\mathbf{a_{2}}} {{O_{2}}^{\mathbf{a_{2}}}}_{\mathbf{b_{2}}} \Psi^{\mathbf{b_{2}}} \overline{\Psi}_{\mathbf{a_{1}}} {{O_{1}}^{\mathbf{a_{1}}}}_{\mathbf{b_{1}}} \Psi^{\mathbf{b_{1}}} \right\rangle &= C^{(0)} + C^{(1)}_{\mathrm{det1}} + C^{(1)}_{\mathrm{det2}} + C^{(1)}_{\mathrm{exch}} + C^{(1)}_{\mathrm{self1}} + C^{(1)}_{\mathrm{self2}}.
\label{eq_expansion_mesonic_2point_function}
\end{align}
The first four contributions are given by the isosymmetric diagram, the two quark-mass-detuning diagrams and an exchange diagram due to the electromagnetic interaction:
\begin{align}
C^{(0)} &= \Bigg\langle
\begin{gathered}
\includegraphics[width=8em]{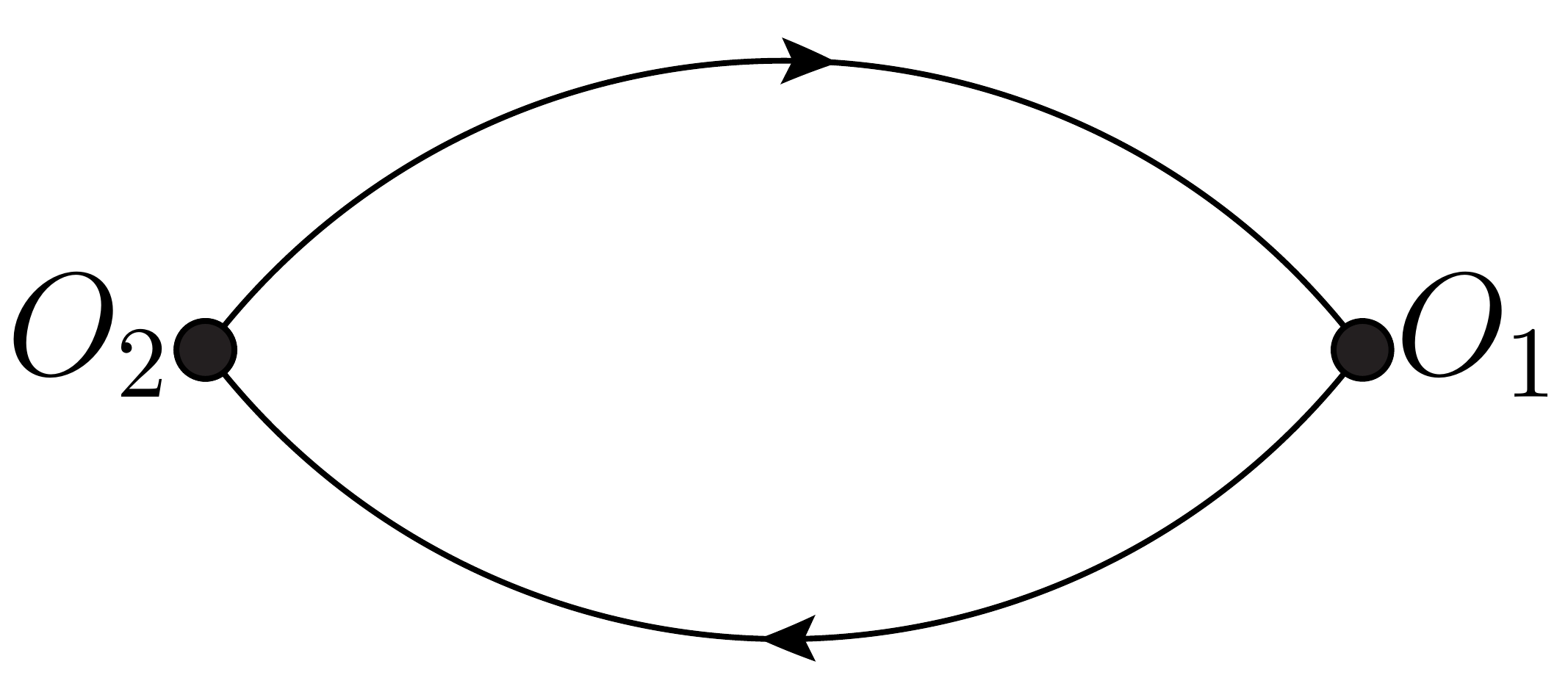}
\end{gathered}
\Bigg\rangle_{\mathrm{g}}^{(0)} & C^{(1)}_{\mathrm{det1}} &= \Bigg\langle
\begin{gathered}
\includegraphics[width=8em]{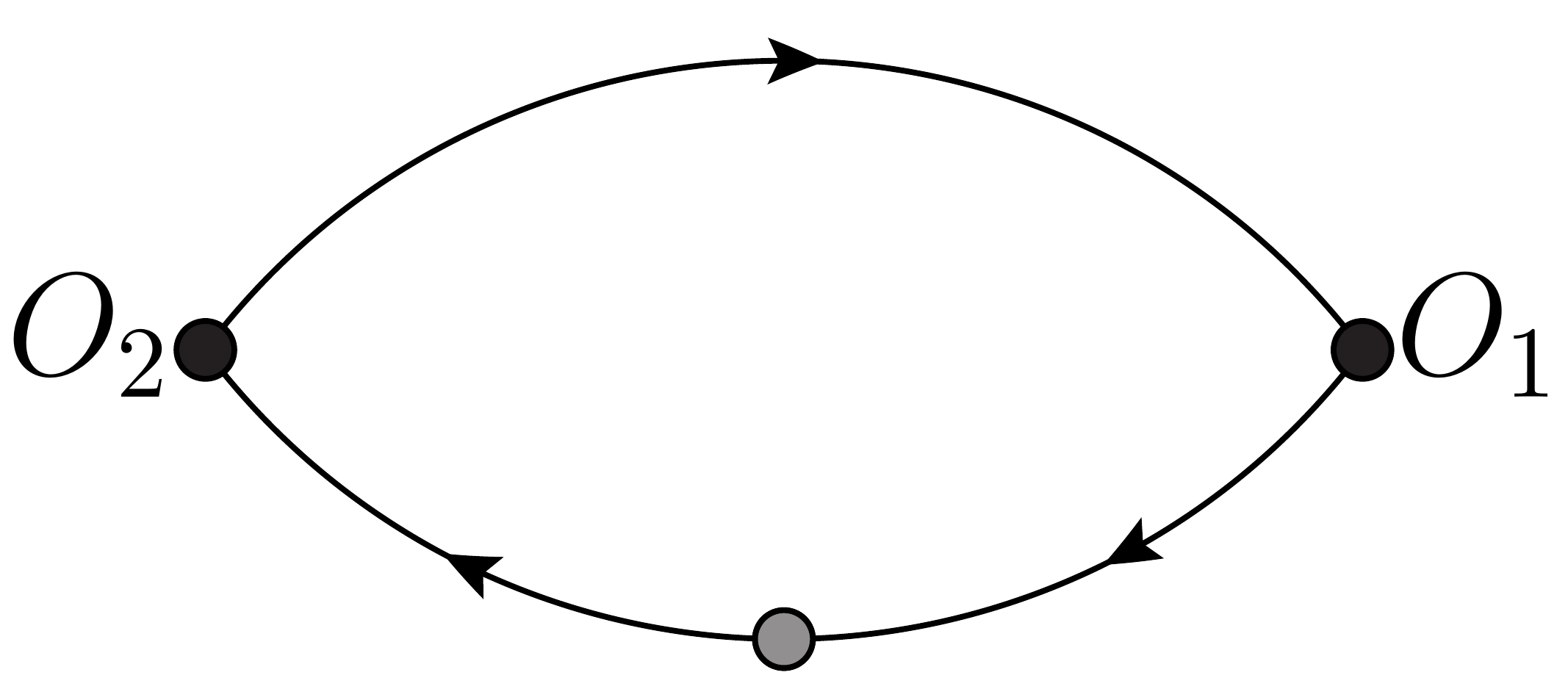}
\end{gathered}
\Bigg\rangle_{\mathrm{g}}^{(0)}
\label{eq_diagrams_expansion_mesonic_2point1} \\
C^{(1)}_{\mathrm{det2}} &= \Bigg\langle
\begin{gathered}
\includegraphics[width=8em]{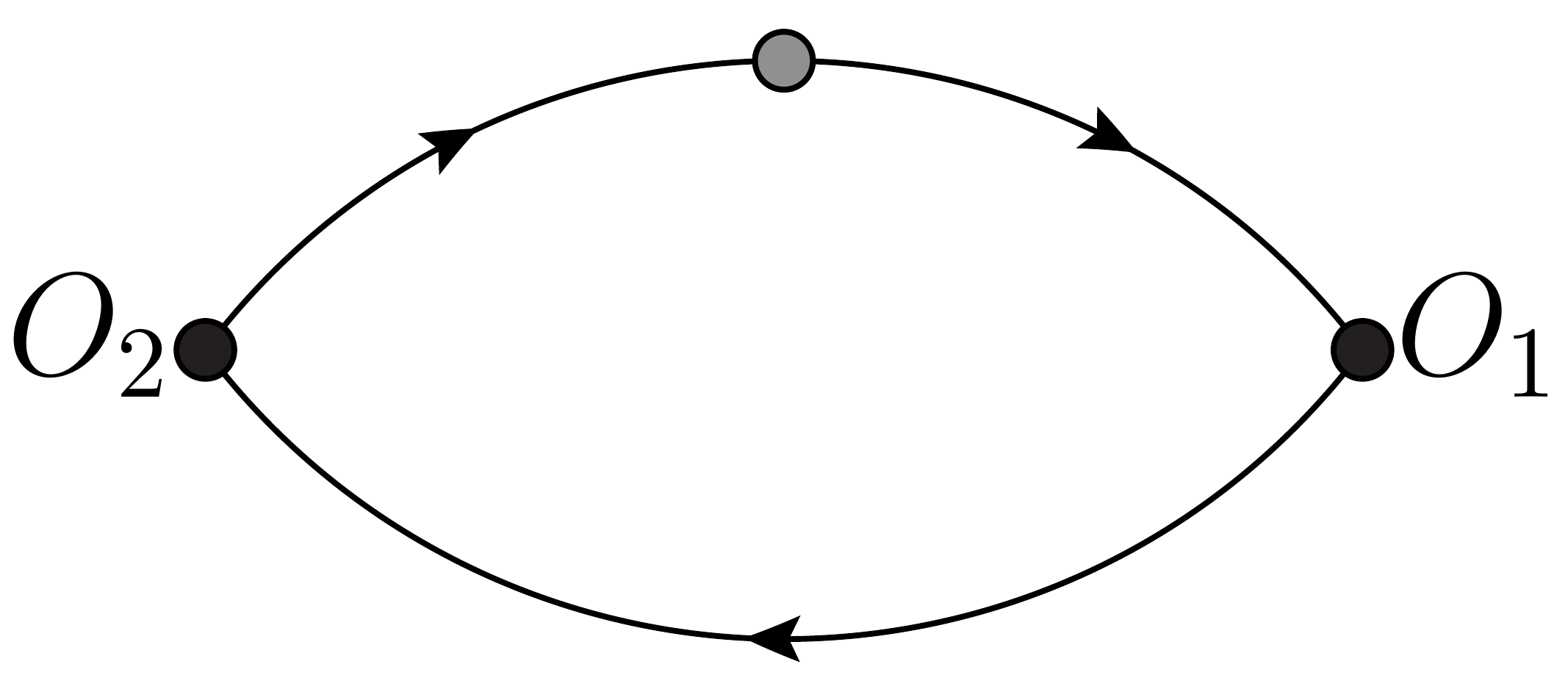}
\end{gathered}
\Bigg\rangle_{\mathrm{g}}^{(0)} & C^{(1)}_{\mathrm{exch}} &= \Bigg\langle
\begin{gathered}
\includegraphics[width=8em]{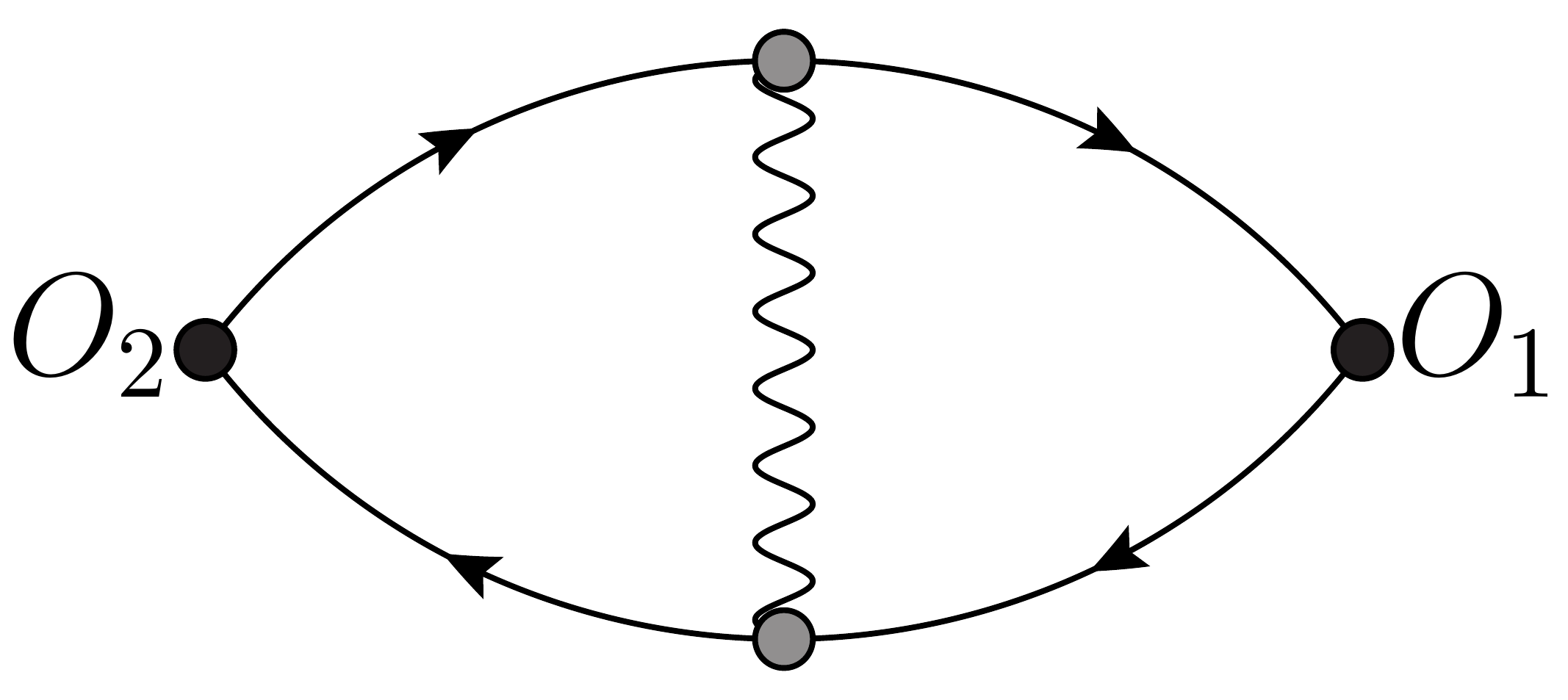}
\end{gathered}
\Bigg\rangle_{\mathrm{g}}^{(0)}. \nonumber
\end{align}
The last two terms are self energy contributions due to the electromagnetic interaction and consist of two diagrams each:
\begin{align}
C^{(1)}_{\mathrm{self1}} &= \Bigg\langle
\begin{gathered}
\includegraphics[width=8em]{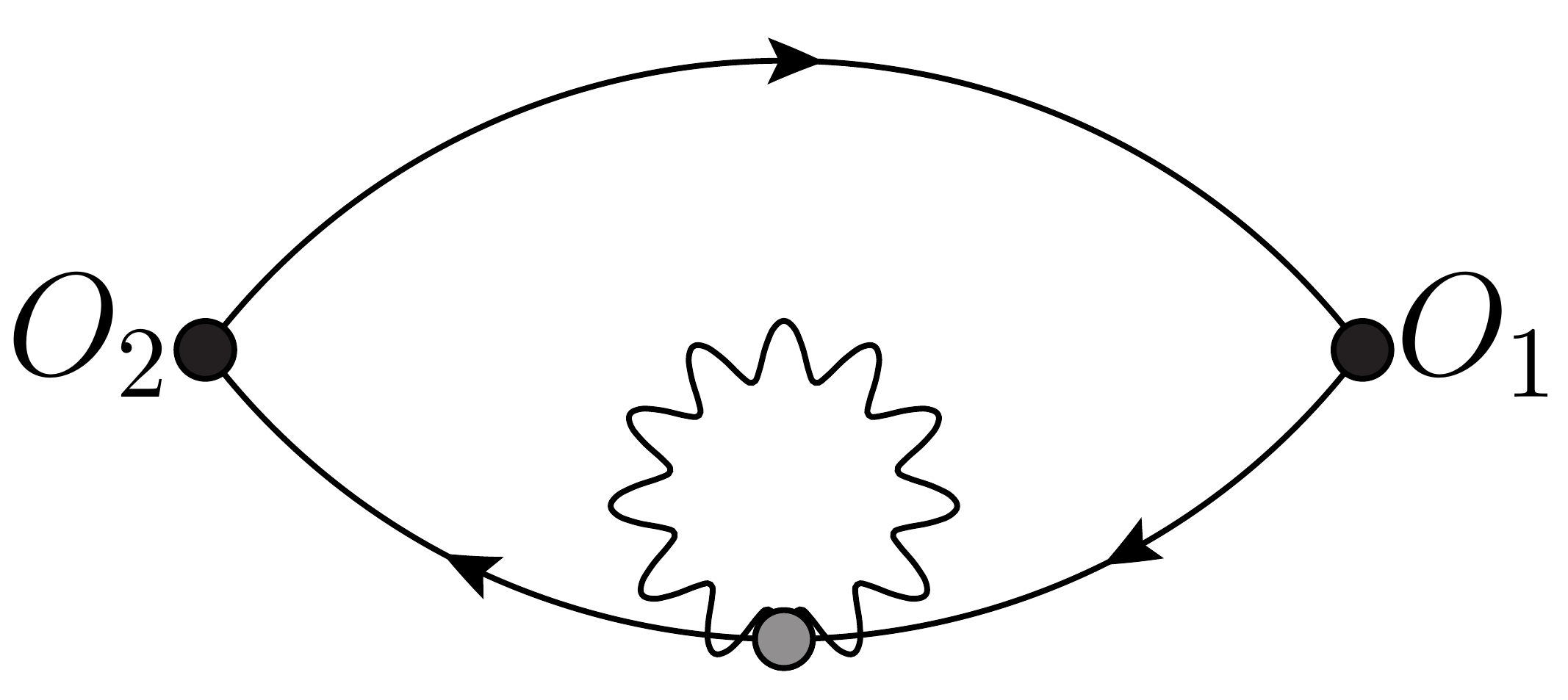}
\end{gathered}
+
\begin{gathered}
\includegraphics[width=8em]{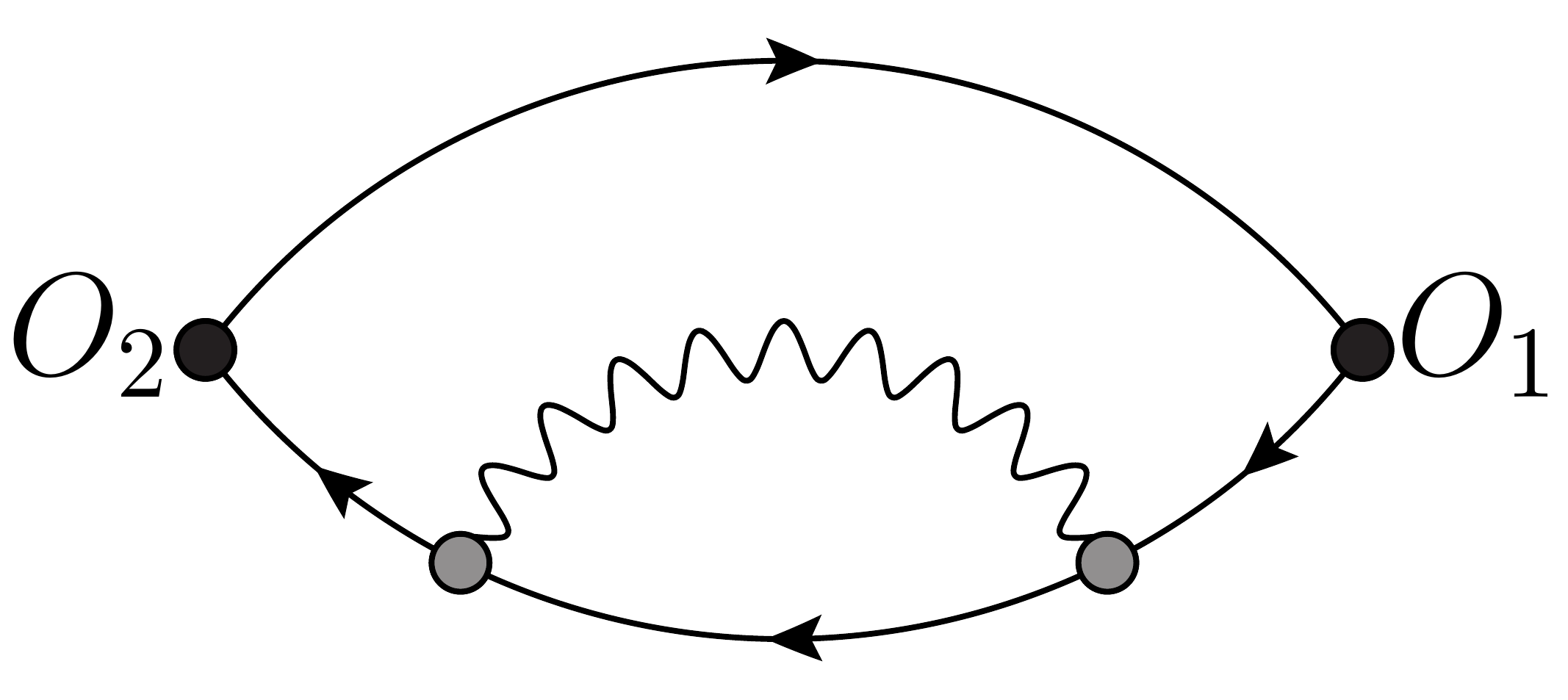}
\end{gathered}
\Bigg\rangle_{\mathrm{g}}^{(0)}
\label{eq_diagrams_expansion_mesonic_2point2} \\
C^{(1)}_{\mathrm{self2}} &= \Bigg\langle
\begin{gathered}
\includegraphics[width=8em]{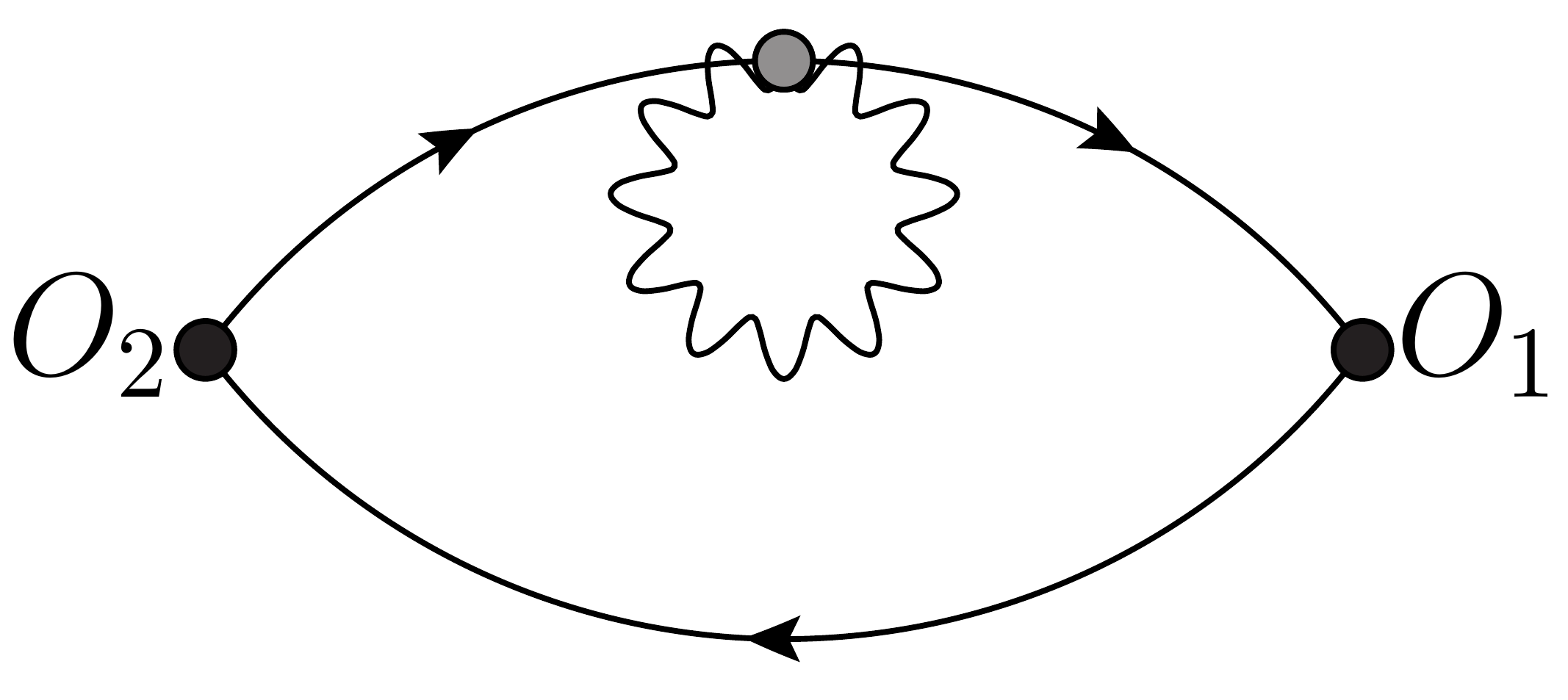}
\end{gathered}
+
\begin{gathered}
\includegraphics[width=8em]{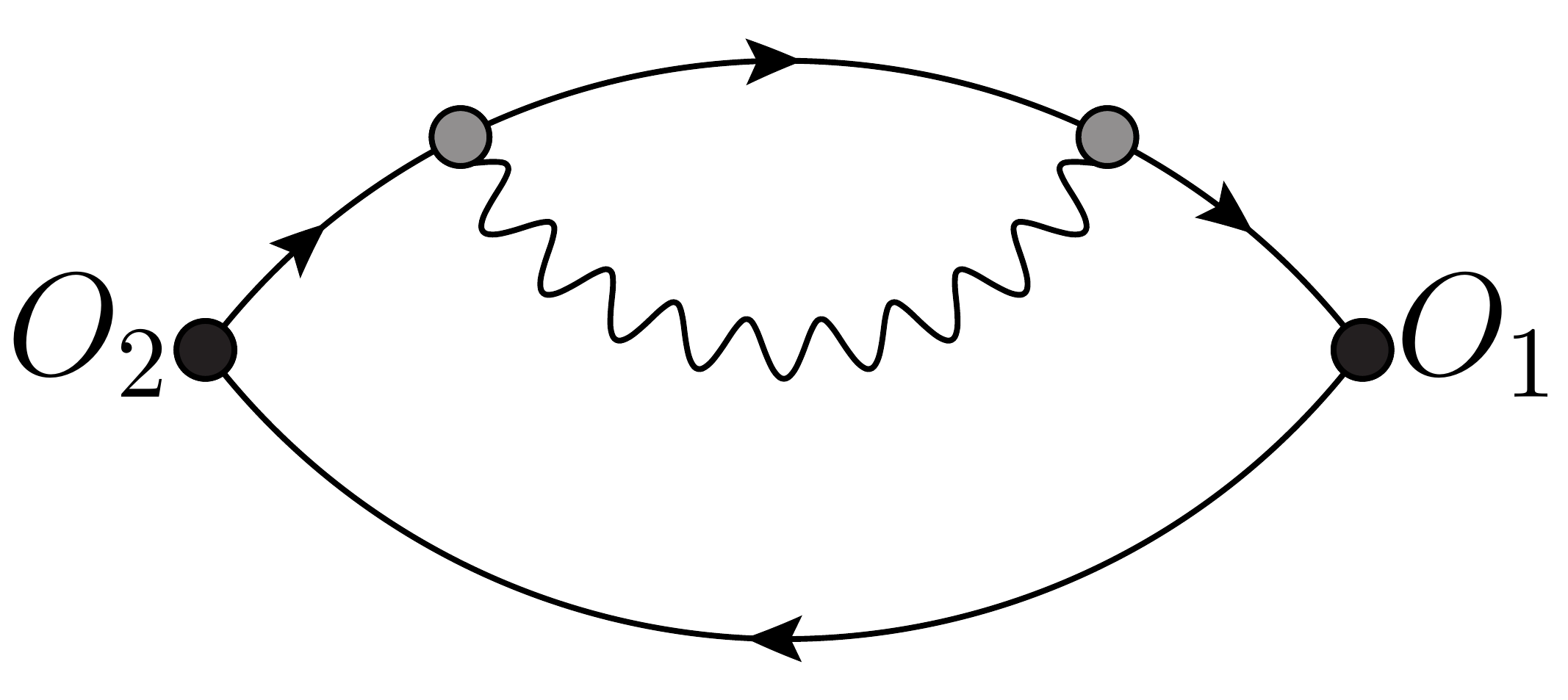}
\end{gathered}
\Bigg\rangle_{\mathrm{g}}^{(0)}. \nonumber
\end{align}

The evaluation of diagrams in which the photon connects two different vertices requires the insertion of an all-to-all photon propagator. We apply a stochastic estimation of the latter~\cite{Tantalo:2013maa, deDivitiis:2013xla, Boyle:2016lbc, Dong:1993pk}. We introduce complex stochastic sources which satisfy the relation $\left\langle J^{\mathbf{c_{2}}} {J^{\dagger}}_{\mathbf{c_{1}}}\right\rangle_{\mathrm{J}} = \delta^{\mathbf{c_{2}}}_{\mathbf{c_{1}}}$. We further define a propagated source $A[J]^{\mathbf{c_{2}}} = {\Sigma^{\mathbf{c_{2}}}}_{\mathbf{c_{1}}} J^{\mathbf{c_{1}}}$. The propagator can now be estimated stochastically by
\begin{align}
\left\langle A[J]^{\mathbf{c_{3}}} {J^{\dagger}}_{\mathbf{c_{1}}} \right\rangle_{\mathrm{J}} &= {\Sigma^{\mathbf{c_{3}}}}_{\mathbf{c_{2}}} \left\langle J^{\mathbf{c_{2}}} {J^{\dagger}}_{\mathbf{c_{1}}}\right\rangle_{\mathrm{J}} = {\Sigma^{\mathbf{c_{3}}}}_{\mathbf{c_{2}}} \delta^{\mathbf{c_{2}}}_{\mathbf{c_{1}}} = {\Sigma^{\mathbf{c_{3}}}}_{\mathbf{c_{1}}}.
\label{eq_stochastic_photon_propagator}
\end{align}
Due to periodic boundary conditions the propagated photon source can be calculated via Fourier transforms. The photon differential operator $\Delta$ becomes diagonal in Fourier space and can therefore be inverted analytically. By means of Eq.~\ref{eq_photon_propagator} the propagated photon source is then calculated by $A[J]^{x_{2}\mu_{2}} = \sum_{p}{{\mathcal{F}^{-1}}^{x_{2}}}_{p} {\widetilde{\Sigma}}^{p\mu_{2}}_{\hphantom{p\mu_{2}}p\mu_{1}} {\mathcal{F}^{p}}_{x_{1}} J^{x_{1}\mu_{1}}$. Whenever both ends of the photon propagator are connected to the same vertex a stochastic estimation can be avoided~\cite{Giusti:2017dmp, Boyle:2017gzv}. The two-quark-two-photon vertex contracted with a photon loop 
can be combined in an effective photon-tadpole vertex:
\begin{align*}
V_{\mathrm{qq\gamma\gamma}}{{}^{\mathbf{a}}}_{\mathbf{bc_{2}c_{1}}}{\Sigma^{\mathbf{c_{1}}}}_{\mathbf{c_{2}}} &=
\begin{gathered}
\includegraphics[width=4em]{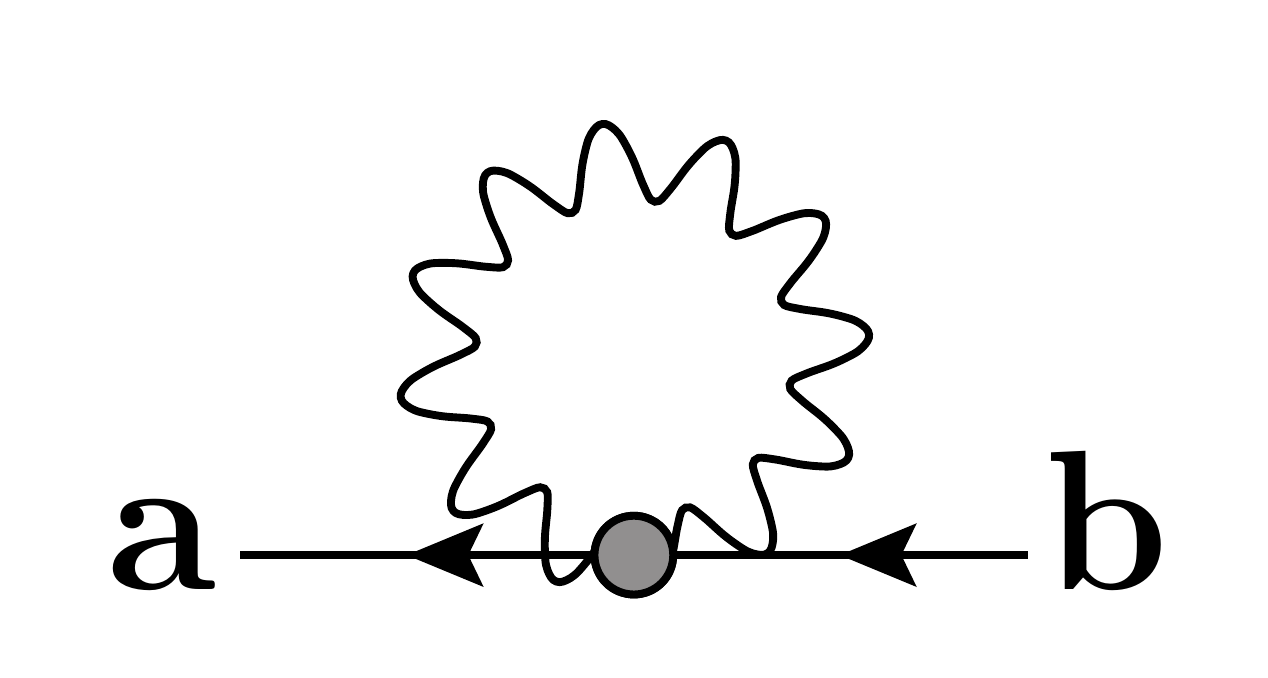}
\end{gathered}
\end{align*}
We can give an analytic expression for the latter vertex as it only depends on diagonal elements of the photon propagator in position space, which are given by ${\Sigma^{x\mu}}_{x\mu} = \sum_{p}{{\mathcal{F}^{-1}}^{x}}_{p} {\widetilde{\Sigma}}^{p\mu}_{\hphantom{p\mu_{2}}p\mu} {\mathcal{F}^{p}}_{x} = \frac{1}{|\Lambda|}\sum_{p}{\widetilde{\Sigma}}^{p\mu}_{\hphantom{p\mu}p\mu}$, where $x$ and $\mu$ are not summed. The result is independent of $x$ as expected due to translational invariance. The photon-loop-contracted vertex applied to a quark field then reads:
\begin{align}
V_{\mathrm{qq\gamma\gamma}}{{}^{xf}}_{\mathbf{bc_{2}c_{1}}}\Sigma^{\mathbf{c_{1}c_{2}}}\Psi^{\mathbf{b}} &= \frac{1}{2}\left(ee^{f}\right)^{2}\sum_{\mu}\left(\left(- \gamma^{\mu} + \mathds{1}\right) U^{x\mu} \Sigma^{x\mu x\mu} \Psi^{x+\hat{\mu},f} + \left(\gamma^{\mu} + \mathds{1}\right) {U^{x-\hat{\mu},\mu}}^{\dagger} \Sigma^{x-\hat{\mu},\mu,x-\hat{\mu},\mu} \Psi^{x-\hat{\mu},f}\right).
\label{eq_contracted_vertices2}
\end{align}

We consider point-like mesonic creation and annihilation operators of the general form $\overline{\Psi}_{\mathbf{a}} {{O}(x)^{\mathbf{a}}}_{\mathbf{b}} \Psi^{\mathbf{b}}= \overline{\Psi}_{x} \Lambda \Gamma \Psi^{x}$,
where $\Lambda$ encodes the quark content and $\Gamma$ the spin structure of the mesons described. The mesonic spectrum can then be extracted from the correlation function
\begin{align*}
C(x^{0}) &= \sum_{\vec{x}\in\Lambda_{123}} \left\langle \overline{\Psi}_{x} \Lambda_{2}\Gamma_{2} \Psi^{x} \, \overline{\Psi}_{0} \Lambda_{1}\Gamma_{1} \Psi^{0} \right\rangle,
\end{align*}
which we expand according to Eq. \ref{eq_expansion_mesonic_2point_function}. As the interpolation operators are point-like we make use of spin, colour and flavour explicit point sources located at the origin $\eta^{\mathbf{a_{2}}}_{\mathbf{a_{1}}} = \delta^{\mathbf{a_{2}}}_{\mathbf{a_{1}}}$, where we fix $x_{1}=0$. In order to evaluate the diagrams defined in Eqs. \ref{eq_diagrams_expansion_mesonic_2point1} and \ref{eq_diagrams_expansion_mesonic_2point2} we define the following sequential propagators:
\begin{align}
S^{(0)}_{V_{\mathrm{qq}}}{{}^{\mathbf{b}}}_{\mathbf{a}} &=
\begin{gathered}
\includegraphics[width=7.5em]{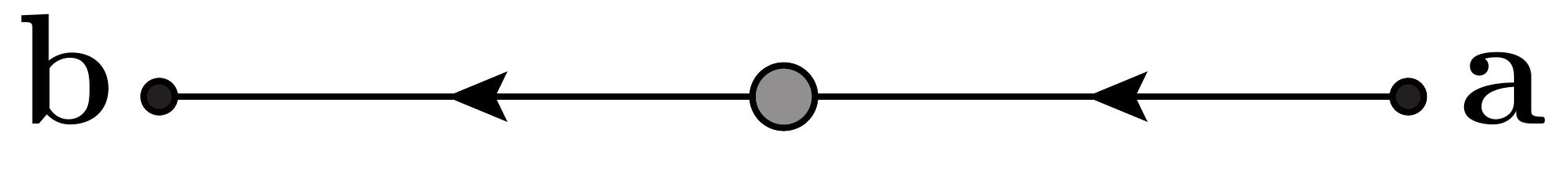}
\end{gathered}
& S^{(0)}_{V_{\mathrm{qq\gamma\gamma}}\Sigma}{{}^{\mathbf{b}}}_{\mathbf{a}} &=
\begin{gathered}
\includegraphics[width=7.5em]{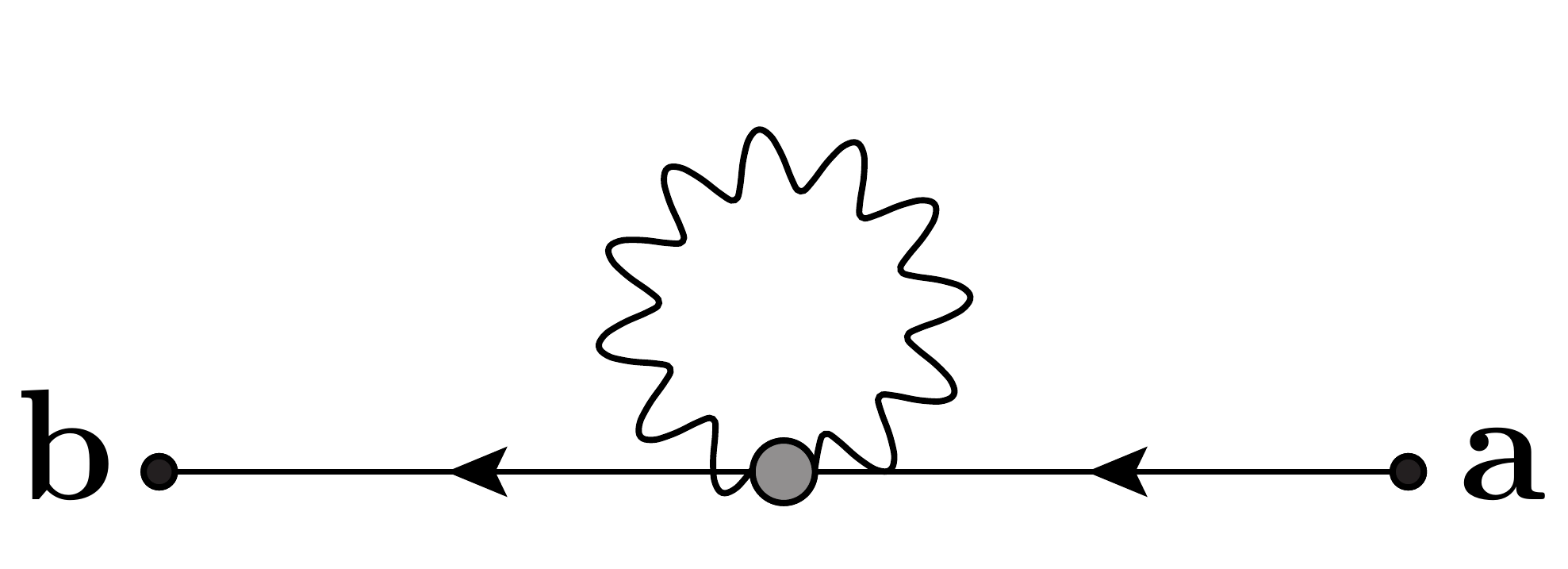}
\end{gathered}
\label{eq_sequential_propagators} \\
S^{(0)}_{V_{\mathrm{qq\gamma}}}[A]{{}^{\mathbf{b}}}_{\mathbf{a}} &=
\begin{gathered}
\includegraphics[width=7.5em]{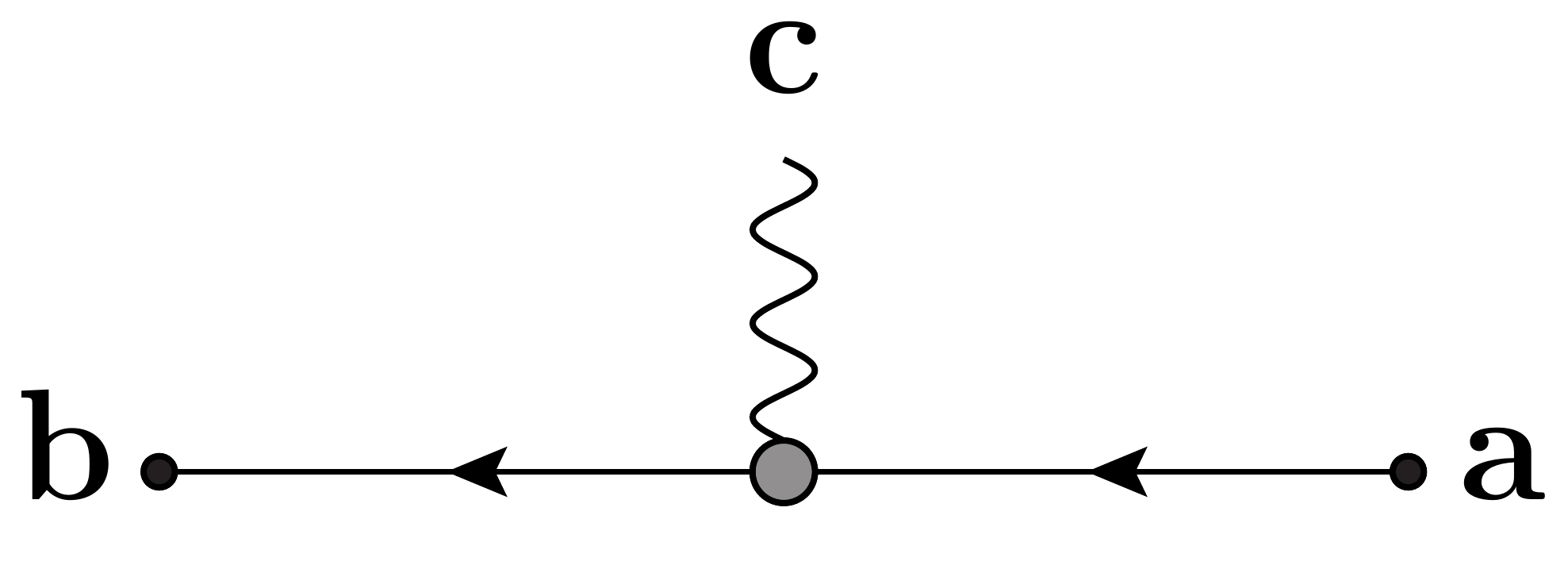}
\end{gathered}
A^{\mathbf{c}} & S^{(0)}_{V_{\mathrm{qq\gamma}}V_{\mathrm{qq\gamma}}}[A_{2},A_{1}]{{}^{\mathbf{b}}}_{\mathbf{a}} &=
\begin{gathered}
\includegraphics[width=10.5em]{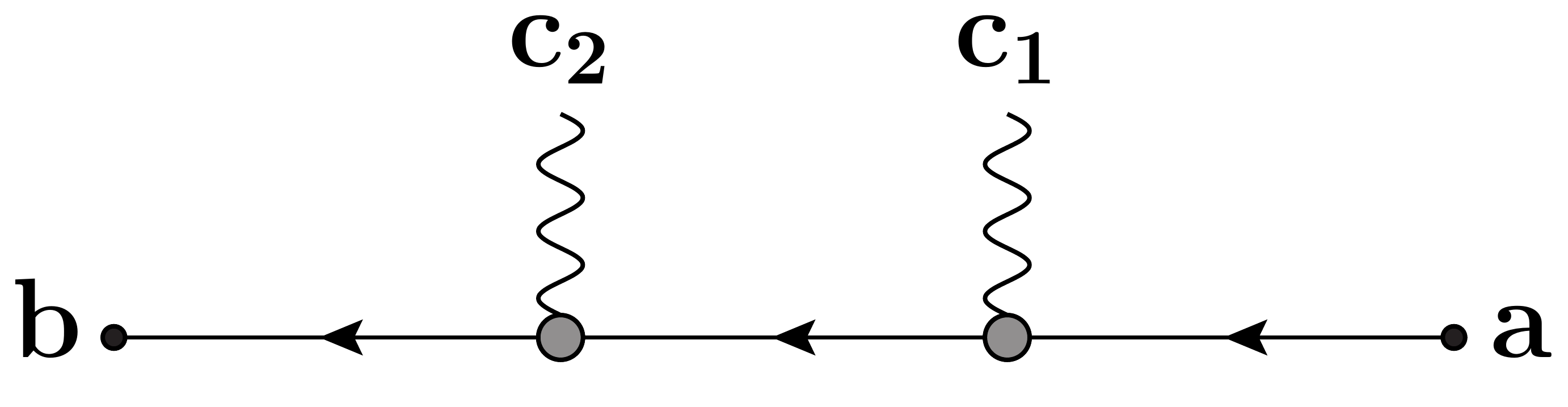}
\end{gathered}
{A_{2}}^{\mathbf{c_{2}}}{A_{1}}^{\mathbf{c_{1}}}. \nonumber
\end{align}
The latter sequential propagators can be calculated iteratively making use of Eqs.~\ref{eq_contracted_vertices1} and \ref{eq_contracted_vertices2} by solving the system
\begin{align*}
D^{(0)}{{}^{\mathbf{a_{2}}}}_{\mathbf{b}}S^{(0)}{{}^{\mathbf{b}}}_{\mathbf{a_{1}}} &= \eta_{\mathbf{a_{1}}}^{\mathbf{a_{2}}} &
D^{(0)}{{}^{\mathbf{a_{2}}}}_{\mathbf{b}}S^{(0)}_{V_{\mathrm{qq}}}{{}^{\mathbf{b}}}_{\mathbf{a_{1}}} &= V_{\mathrm{qq}}{{}^{\mathbf{a_{2}}}}_{\mathbf{b}}S^{(0)}{{}^{\mathbf{b}}}_{\mathbf{a_{1}}} \\
D^{(0)}{{}^{\mathbf{a_{2}}}}_{\mathbf{b}}S^{(0)}_{V_{\mathrm{qq}\gamma\gamma}\Sigma}{{}^{\mathbf{b}}}_{\mathbf{a_{1}}} &= V_{\mathrm{qq\gamma\gamma}}{{}^{\mathbf{a}}}_{\mathbf{bc_{2}c_{1}}}\Sigma^{\mathbf{c_{1}c_{2}}}S^{(0)}{{}^{\mathbf{b}}}_{\mathbf{a_{1}}} &
D^{(0)}{{}^{\mathbf{a_{2}}}}_{\mathbf{b}}S^{(0)}_{V_{\mathrm{qq\gamma}}}[A]{{}^{\mathbf{b}}}_{\mathbf{a_{1}}} &= V_{\mathrm{qq\gamma}}{{}^{\mathbf{a_{2}}}}_{\mathbf{bc}}S^{(0)}{{}^{\mathbf{b}}}_{\mathbf{a_{1}}}A^{\mathbf{c}} \\
D^{(0)}{{}^{\mathbf{a_{2}}}}_{\mathbf{b}}S^{(0)}_{V_{\mathrm{qq\gamma}}V_{\mathrm{qq\gamma}}}[A_{2},A_{1}]{{}^{\mathbf{b}}}_{\mathbf{a_{1}}} &= V_{\mathrm{qq\gamma}}{{}^{\mathbf{a_{2}}}}_{\mathbf{bc}}S^{(0)}_{V_{\mathrm{qq\gamma}}}[A_{1}]{{}^{\mathbf{b}}}_{\mathbf{a_{1}}}{A_{2}}^{\mathbf{c}}.
\end{align*}
The isosymmetric contribution to the 2-point correlation function is calculated by
\begin{align*}
C^{(0)}(x^{0}) & = - \sum_{f_{1},f_{2}}\Lambda_{2,f_{2}f_{1}}\sum_{\vec{x}\in\Lambda_{123}}\left\langle \Tr\left(\gamma_{5}\,S^{(0)}{{{}^{xf_{2}}}_{0f_{2}}}^{\dagger}\,\gamma_{5}\Gamma_{2}\,S^{(0)}{{}^{xf_{1}}}_{0f_{1}}\,\Gamma_{1}\right)\right\rangle_{\mathrm{g}}^{(0)}\Lambda_{1,f_{2}f_{1}},
\end{align*}
where we made use of the $\gamma_{5}$-hermiticity of the isosymmetric propagator $S^{(0)}$. As the expansion parameters depend on the quark flavours we split up the first-order contribution
\begin{align*}
C^{(1)}_{diag}(x^{0}) &= \sum_{f_{1},f_{2}}\Lambda_{2,f_{2}f_{1}}\,C^{(1)}_{diag,f_{2}f_{1}}(x^{0})\,\Lambda_{1,f_{2}f_{1}},
\end{align*}
where $diag\in\{\mathrm{det1},\mathrm{det2},\mathrm{exch},\mathrm{self1},\mathrm{self2}\}$, and $f_{1}$ and $f_{2}$ denote the forward and backward running quark flavours, respectively. For the diagrams $\mathrm{det1}$, $\mathrm{exch}$ and $\mathrm{self2}$ we give the explicit form of the correlation function in terms of sequential propagators defined in Eq. \ref{eq_sequential_propagators} and of the stochastic estimate of the photon propagator defined in Eq. \ref{eq_stochastic_photon_propagator}:
\begin{align*}
C^{(1)}_{\mathrm{det1},f_{2}f_{1}}(x^{0}) &= \sum_{\vec{x}\in\Lambda_{123}}\left\langle\Tr\left(\gamma_{5}\,S^{(0)}{{{}^{xf_{2}}}_{0f_{2}}}^{\dagger}\,\gamma_{5}\Gamma_{2}\,S^{(0)}_{V_{\mathrm{qq}}}{{}^{xf_{1}}}_{0f_{1}}\,\Gamma_{1}\right)\right\rangle_{\mathrm{g}}^{(0)} \\
C^{(1)}_{\mathrm{exch},f_{2}f_{1}}(x^{0}) &= - \sum_{\vec{x}\in\Lambda_{123}}\left\langle\left\langle\Tr\left(\gamma_{5}\,S^{(0)}_{V_{\mathrm{qq\gamma}}}[A[J]^{\dagger}]{{{}^{xf_{2}}}_{0f_{2}}}^{\dagger}\,\gamma_{5}\Gamma_{2}\,S^{(0)}_{V_{\mathrm{qq\gamma}}}[J^{\dagger}]{{}^{xf_{1}}}_{0f_{1}}\,\Gamma_{1}\right)\right\rangle_{\mathrm{J}}\right\rangle^{(0)}_{\mathrm{g}} \\
C^{(1)}_{\mathrm{self2},f_{2}f_{1}}(x^{0}) &= \frac{1}{2} \sum_{\vec{x}\in\Lambda_{123}}\left\langle\Tr\left(\gamma_{5}\,S^{(0)}_{V_{\mathrm{qq\gamma\gamma}}\Sigma}{{{}^{xf_{2}}}_{0f_{2}}}^{\dagger}\,\gamma_{5}\Gamma_{2}\,S^{(0)}{{}^{xf_{1}}}_{0f_{1}}\,\Gamma_{1}\right)\right\rangle^{(0)}_{\mathrm{g}} \\
&\hphantom{=} - \sum_{\vec{x}\in\Lambda_{123}}\left\langle\left\langle\Tr\left(\gamma_{5}\,S^{(0)}_{V_{\mathrm{qq\gamma}}V_{\mathrm{qq\gamma}}}[A[J],J^{\dagger}]{{{}^{xf_{2}}}_{0f_{2}}}^{\dagger}\,\gamma_{5}\Gamma_{2}\,S^{(0)}{{}^{xf_{1}}}_{0f_{1}}\,\Gamma_{1}\right)\right\rangle_{\mathrm{J}}\right\rangle^{(0)}_{\mathrm{g}}.
\end{align*}
In addition, we made use of the $\gamma_{5}$-hermiticity of the vertices contracted with photon fields~\cite{deDivitiis:2013xla}, which can be derived from the expansion of the $\gamma_{5}$-hermitian Dirac operator $D$. The explicit expressions for the correlations functions $\mathrm{det2}$ and $\mathrm{self1}$ can be derived from the examples given above.

The asymptotic behaviour of mesonic 2-point correlation functions in the limit $0\ll x^{0}\ll T$ is given by $C(x^{0}) = a\cosh\left(m\left(\frac{T}{2}-x^{0}\right)\right)$. Expanding the parameters $m$ and $a$ in $\Delta m_{f}=m_{f}-m^{(0)}_{f}$ with $f\in \{\mathrm{u},\mathrm{d},\mathrm{s}\,\ldots\}$ and $e^{2}$ yields $m=m^{(0)}+m^{(1)}+...$ and $a=a^{(0)}+a^{(1)}+...$ . The parameters $m^{(0)}$ and $a^{(0)}$ are extracted from the zeroth-order contribution to the correlation function
\begin{align}
C^{(0)}(x^{0}) &= a^{(0)}\cosh\left(m^{(0)}\left(\frac{T}{2}-x^{0}\right)\right).
\label{eq_asymptotic_correlation_function_zeroth_order}
\end{align}
In order to obtain the first-order contributions $m^{(1)}$ and $a^{(1)}$ we build ratios of first-order correlation functions and the zeroth-order correlation function ~\cite{Sanfilippo:2011zz, deDivitiis:2011eh, Tantalo:2013maa, deDivitiis:2013xla}
\begin{align}
\frac{C^{(1)}_{diag,f_{1}f_{2}}(x^{0})}{C^{(0)}(x^{0})} &= \frac{a^{(1)}_{diag,f_{1}f_{2}}}{a^{(0)}}+m^{(1)}_{diag,f_{1}f_{2}}\left(\frac{T}{2}-x^{0}\right) \tanh\left(m^{(0)}\left(\frac{T}{2}-x^{0}\right)\right).
\label{eq_asymptotic_correlation_function_first_order}
\end{align}
$a^{(1)}_{diag,f_{1}f_{2}}$ and $m^{(1)}_{diag,f_{1}f_{2}}$ are extracted by fitting making use of the zeroth-order contributions determined beforehand. The final ground state mass to first-order is then given by
\begin{align}
m &= m^{(0)}+\sum_{diag}\sum_{f_{1},f_{2}}\Lambda_{2,f_{2}f_{1}} m^{(1)}_{diag,f_{1}f_{2}}\Lambda_{1,f_{1}f_{2}}.
\label{eq_mass}
\end{align}

\begin{table}[h]
  \centering
  \caption{Lattice parameters of E5~\cite{Capitani:2015sba}}
\begin{tabular}{|l|l|l|l|l|l|l|}
\hline
$\mathrm{T} \times \mathrm{L}^3$ & $\beta$ & $\kappa_{\mathrm{l}}$ & $\kappa_{\mathrm{s}}$ & $a[\mathrm{fm}]$ & $m_{\pi}[\mathrm{MeV}]$ \\ \hline
$64 \times 32^{3}$ & $5.30$ & $0.13625$ & $0.135802302$ & $0.0658$ & $437$ \\ \hline
\end{tabular}
\label{table_lattice_parameters_e5}
\end{table}

The simulation code is based on the OpenQCD~\cite{Luscher:} and QDP++~\cite{Edwards:2004sx} frameworks. For the Fourier transform of the stochastic photon sources we made use of the FFTW3 library~\cite{FFTW05}. A first analysis was performed on the CLS ensemble E5 with lattice parameters given in Table~\ref{table_lattice_parameters_e5} comprising 1000 gauge configurations. Per gauge configuration 2 randomly positioned quark point sources were used as well as 8 individual stochastic photon sources per quark source. We made use of real $\mathds{Z}_{2}$ photon sources, where $J^{x\mu} \in \{-1,1\}$~\cite{Dong:1993pk}. Before estimating the statistical error via bootstrap resampling we apply a blocking to all 16 measurements performed on the same gauge configuration. Fits were performed in the interval $[16,30]$. A systematic fit error was estimated via the dispersion of the fit results in the intervals $[16\pm\{0,1,2\},30\pm\{0,1,2\}]$.

\begin{figure}
\centering
\begin{subfigure}[c]{0.49\textwidth}
\includegraphics[width=\textwidth]{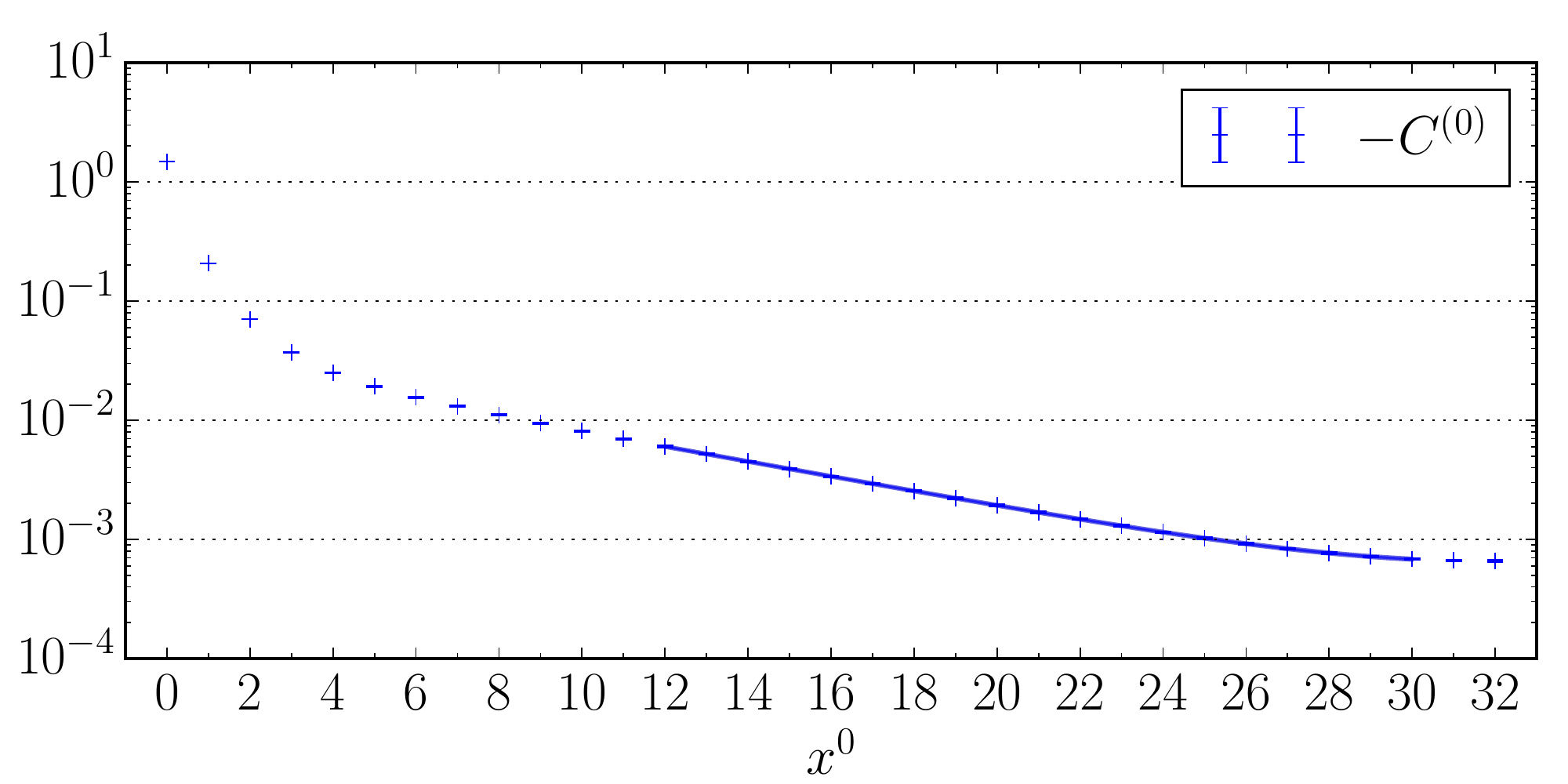}
\subcaption{Isosymmetric contribution}
\label{subfig_pion+_isosymmetric_contribution}
\end{subfigure}
\begin{subfigure}[c]{0.49\textwidth}
\includegraphics[width=\textwidth]{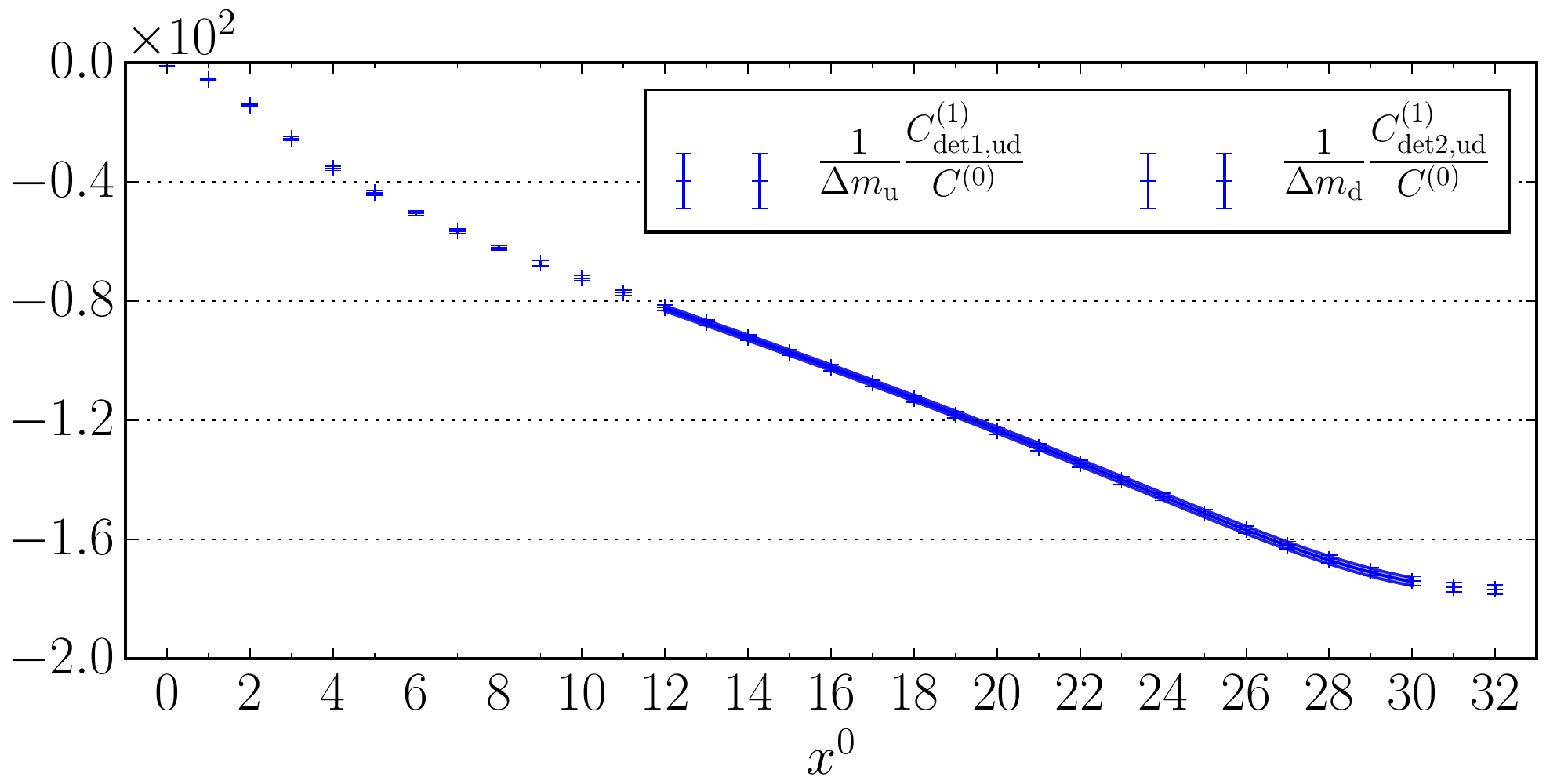}
\subcaption{Normalised mass detuning contribution}
\label{subfig_pion+_mass_detuning_contribution}
\end{subfigure}
\begin{subfigure}[c]{0.49\textwidth}
\includegraphics[width=\textwidth]{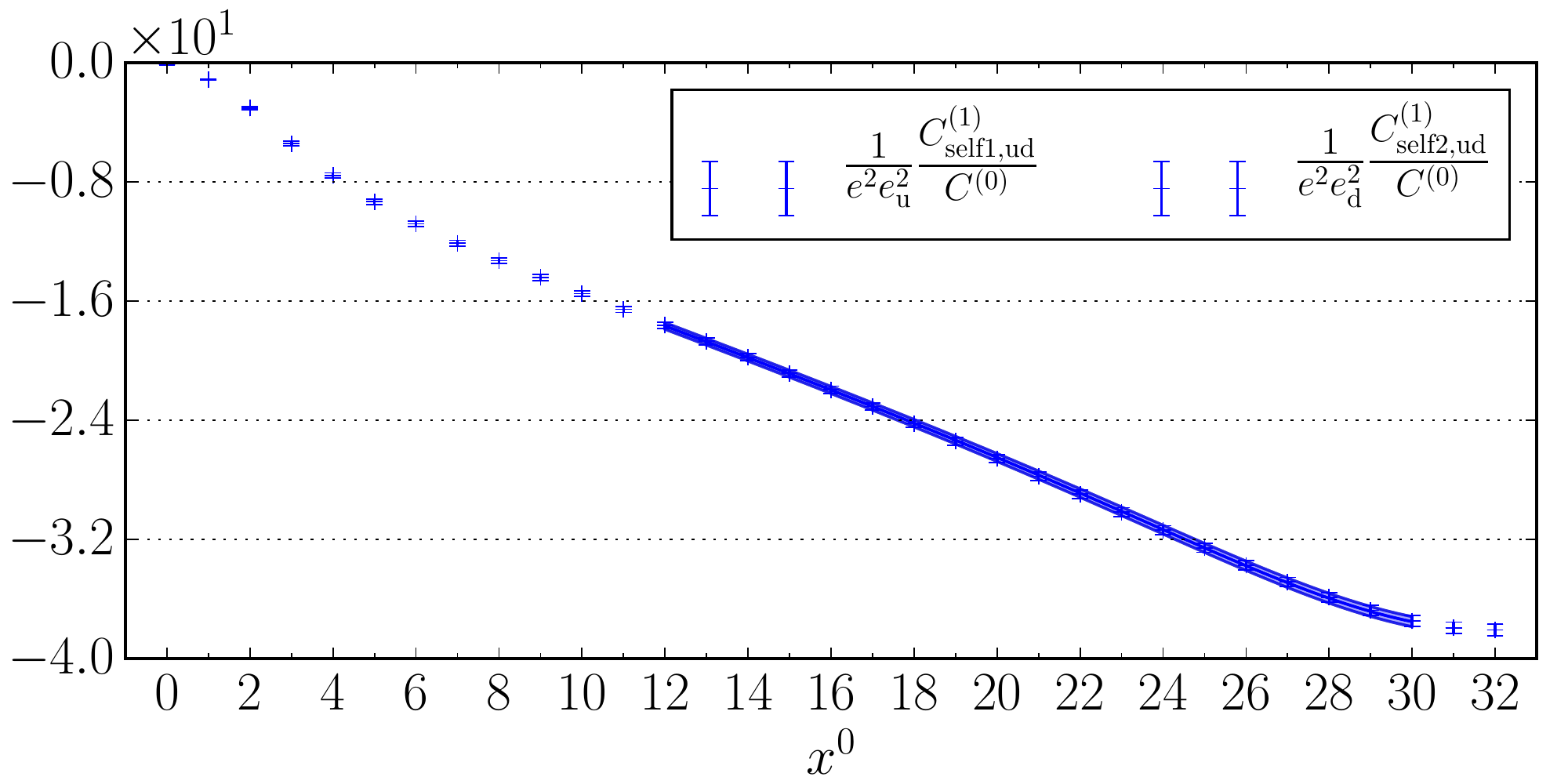}
\subcaption{Normalised self energy contribution}
\label{subfig_pion+_self_energy_contribution}
\end{subfigure}
\begin{subfigure}[c]{0.49\textwidth}
\includegraphics[width=\textwidth]{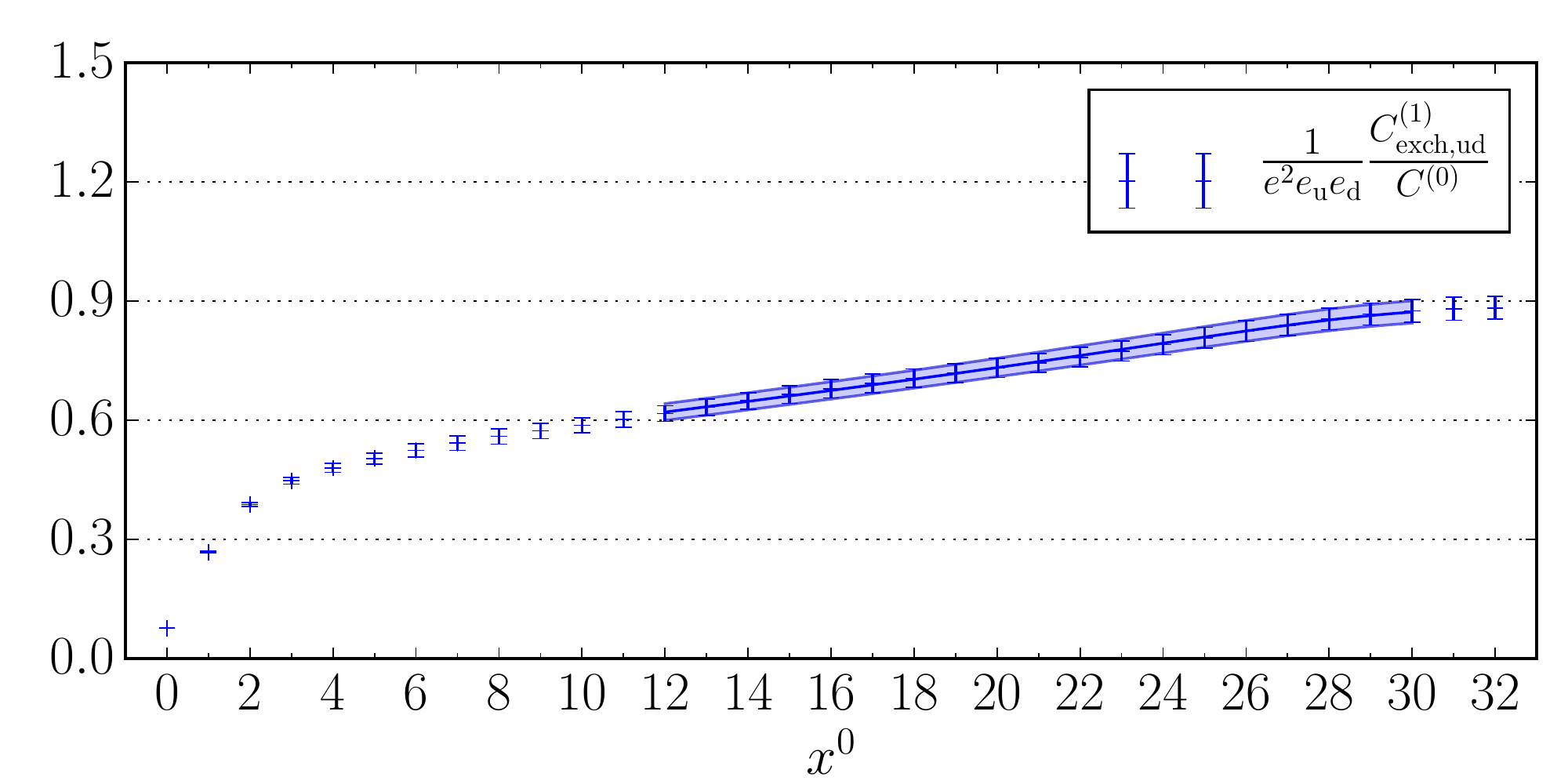}
\subcaption{Normalised photon exchange contribution}
\label{subfig_pion+_photon_exchange_contribution}
\end{subfigure}
\caption{Correlation functions for the $\pi^{+}$ meson defined in Eqs.~ \ref{eq_diagrams_expansion_mesonic_2point1} and \ref{eq_diagrams_expansion_mesonic_2point2} in lattice units. The fit in Fig.~\ref{subfig_pion+_isosymmetric_contribution} is performed with respect to Eq.~\ref{eq_asymptotic_correlation_function_zeroth_order} and in Figs.~\ref{subfig_pion+_mass_detuning_contribution},~\ref{subfig_pion+_self_energy_contribution} and~\ref{subfig_pion+_photon_exchange_contribution} with respect to Eq.~\ref{eq_asymptotic_correlation_function_first_order}.}
\label{fig_pion+}
\end{figure}

In Fig.~\ref{fig_pion+} we show correlation functions in the $\pi^{+}$-channel. Correlation functions for the $\pi^{0}$, $K^{+}$ and $K^{0}$ mesons look comparable. In our simulation we have determined pseudo-scalar meson masses for the kaon and pion sector according to Eq.~\ref{eq_mass}. The average masses and mass splittings of isospin doublets $K$ and $\pi$ in lattice units read
\begin{align*}
a\left(m_{\pi^{+}}-m_{\pi^{0}}\right)&= 0.000594(43)(7) \\
\frac{a}{2}\left(m_{\pi^{+}}+m_{\pi^{0}}\right)&= 0.14536(60)(11) + a\left(\Delta m_{\mathrm{u}}+\Delta m_{\mathrm{d}}\right) \cdot 4.718(50)(14) + 0.05204(62)(17)\\
a\left(m_{K^{+}}-m_{K^{0}}\right)&= a\left(\Delta m_{\mathrm{u}}-\Delta m_{\mathrm{d}}\right) \cdot 3.471(32)(9) + 0.02337(23)(6)\\
\frac{a}{2}\left(m_{K^{+}}+m_{K^{0}}\right)&= 0.19349(47)(10) + a\left(\Delta m_{\mathrm{u}}+\Delta m_{\mathrm{d}}\right) \cdot 3.471(32)(9) \\
&\hphantom{=}+a\Delta m_{\mathrm{s}} \cdot 3.474(13)(04) + 0.02696(22)(6),
\end{align*}
where the first term in the averages is given by the isosymmetric contribution and the last term by the electromagnetic contribution respectively. The first error accounts for the statistical accuracy and the second error for the systematic uncertainty in the choice of the fit interval. The pion mass splitting is a pure first-order electro-magnetic effect~\cite{deDivitiis:2013xla}, such that we can use the isosymmetric scale from Table~\ref{table_lattice_parameters_e5} to convert to physical units:
\begin{align*}
m_{\pi^{+}}-m_{\pi^{0}} &= 1.781(130)(22)\,\mathrm{MeV}.
\end{align*}
This value differs significantly from the experimentally determined value $4.5936(5)\,\mathrm{MeV}$~\cite{Patrignani:2016xqp}, which is not unexpected as we have not taken into account finite volume corrections, which are large due to the long-range nature of the electromagnetic interaction, especially for large unphysical pion masses ($437\,\mathrm{MeV}$ in this simulation) and lead to a strongly increased mass splitting~\cite{Borsanyi:2014jba, Giusti:2017dmp, Boyle:2017gzv}. Furthermore, we have not performed a continuum extrapolation and we have neglected quark-disconnected diagrams as well as vacuum diagrams, which presumably only give small contributions.

\section{Conclusions and Outlook}
\label{sec_conlusions}

For the investigation of leading order isospin breaking effects we gave a set of Feynman rules based on $O(a)$ improved Wilson fermions. As a first application we calculated leading isospin breaking corrections to pseudo-scalar meson masses on one ensemble. An independent calculation with lower statistics using a photon propagator in Coulomb gauge lead to compatible results. We plan to extend our investigations to other observables such as the anomalous magnetic moment of the muon on the open boundary gauge ensembles created in the CLS effort.

\begin{acknowledgement}

We  are  grateful  to  our  colleagues  within  the  CLS  initiative  for  sharing ensembles.  Our calculations were performed on the HPC Cluster ``Clover`` at the Helmholtz Institute Mainz. We thank Georg von Hippel for helpful discussions. Andreas Risch is a recipient of a fellowship through GRK
Symmetry Breaking (DFG/GRK 1581).

\end{acknowledgement}

\bibliography{lattice2017}

\end{document}